# Bulk Photovoltaic Effect in Two-Dimensional Perovskite Oxides

Chunmei Zhang[1,2,3] Jian Zhou[4*], Liang Si[1,2,3*]

[1]School of Physics, Northwest University, Xi'an 710127, China
[2]Shaanxi Key Laboratory for Theoretical Physics Frontiers, Xi'an 710127, China
[3]Fundamental Discipline Research Center for Quantum Science and Technology of Shaanxi Province, Xi'an 710127, China
[4]Center for Alloy Innovation and Design, State Key Laboratory for Mechanical Behavior of Materials, Xi'an Jiaotong University, Xi'an, 710049, China

[*]Corresponding authors: jianzhou@xjtu.edu.cn; siliang@nwu.edu.cn

## Abstract

Perovskite oxides $ABO_3$ host a rich interplay of charge, spin, lattice, and orbital degrees of freedom, giving rise to diverse quantum phenomena. In low-dimensional $ABO_3$, reduced symmetry can induce exotic quantum effects such as the two-dimensional electron gas and unconventional superconductivity. Using first-principles density functional theory, tight-binding modeling, and symmetry analysis, we show that ultrathin two-dimensional (2D) $ABO_3$ films—exemplified by $SrTiO_3$—naturally break inversion symmetry, producing a spontaneous out-of-plane bulk photovoltaic (BPV) effect. This differs from previous studies on in-plane BPV current signals and is more applicable and experimentally detectable. Such an effect is highly tunable via thickness, strain, surface termination, crystallographic orientation, and Moiré twisting. These findings are broadly applicable to a wide range of 2D perovskite and other layer-resolved oxides.

## Introduction

Perovskite oxides with the general formula $ABO_3$ represent a foundational but important family of quantum materials [1], which host a wide range of emergent quantum phenomena [2], including unconventional superconductivity [3], ferroelectricity [4], multiferroicity [5], and Mott-insulator transition [2]. The remarkable tunability of these properties—via strain, dimensional confinement, or chemical substitution—makes perovskites highly versatile platforms for both fundamental research and technological applications. Specifically, in reduced dimensions, such as ultrathin films, surfaces, or interfaces, perovskite systems often exhibit properties absent in their bulk counterparts [2]. Reduced symmetry operators in these low-dimensional states, such as the absence of an inversion center, give rise to novel quantum phenomena, including magnetic skyrmions [6], Rashba-type spin splitting [7], and the bulk photovoltaic (BPV) effect [8]. In particular, the BPV effect is a second-order nonlinear optical response. It usually enables the generation of a steady-state closed circuit current or open circuit voltage without the formation of *p–n* junctions and could potentially overcome the well-known Schokley-Queisser limit. According to the microscopic picture, the BPV effect is sensitive to the Bloch band quantum geometry and is susceptible to lattice symmetry [9-12]. The intrinsic mechanisms are partitioned as the shift current and injection current [13,14]. The shift current originates from the Wannier function center evolution via interband electron transition. The injection current is an intrinsic mechanism of ballistic current, arising from inequivalent carrier generation at $\boldsymbol{k}$ and $-\boldsymbol{k}$. In nonmagnetic systems, they are induced by linearly polarized light (LPL) and circularly polarized light (CPL), respectively [15]. Both of them are rooted in topological phases of electron wavefunctions, such as shift vectors and Berry connection [16-18].

The BPV mechanism was first proposed in perovskites, which provide an ideal platform for exploring asymmetry-driven optical responses [12,19]. However, most bulk $ABO_3$ perovskite oxides, such as bulk $SrTiO_3$ (STO) or $SrRuO_3$, are centrosymmetric in a wide range of external pressure and environmental temperature and thus lack an intrinsic BPV effect, limiting their utility in this context. Nonetheless, $ABO_3$ perovskites remain promising candidates for BPV engineering due to several advantageous features. First, they offer a wide range of band gaps (0.6-4.7 eV), which play a critical role in determining optical activity. Second, low-dimensional forms—such as thin films and interfaces—can be synthesized via epitaxial techniques such as pulsed laser deposition (PLD) and molecular beam epitaxy (MBE) [20-26]. Third, surface and termination engineering are well-established in both the Ruddlesden-Popper (RP) series ($A_{n+1}B_nO_{3n+1}$) and their infinite-layer counterpart $ABO_3$, offering additional means to control the lattice symmetry. A very recent study reported a universal method for in situ control of stoichiometry and termination of epitaxial perovskite films, exploiting dynamic B-site layer inversion observed in in situ electron diffraction [27]. This method works for a wide range of perovskite $ABO_3$ phases, indicating that precise termination control is becoming a general capability in oxide thin-film growth [28,29]. Finally, recent advances introduce a new degree of freedom, twisted stacking with arbitrary angles in layered perovskites. This has been experimentally realized in $BaTiO_3$ (BTO) [30,31] and STO [32-35]. In such twisted heterostructures, their lattices form square-domain Moiré superlattices and are distinct from the commonly studied triangular or honeycomb lattices, such as those in twisted bilayer graphene [36-40]. Moreover, with the development of water-soluble sacrificial substrates and thin-film exfoliation techniques [22,41], the fabrication of freestanding two dimensional (2D) films and even Moiré-twisted oxide structures [33,42-45] has become feasible. Hence, both theoretical and experimental efforts are warranted to explore low-dimensional symmetry breaking or Moiré superlattices in $ABO_3$ and thereby tune their BPV effect, including the magnitude and frequency responses.

In this work, using first-principles density functional theory (DFT), tight-binding (TB) model, and symmetry analysis, we investigate the BPV effect in ultrathin, stoichiometric 2D $ABO_3$ films that intrinsically break inversion symmetry even without structural distortions (such as octahedral rotations) or chemical asymmetry. Taking STO as a model system, we show that this symmetry breaking gives rise to a spontaneous BPV response. Group theory analysis reveals that only out-of-plane photovoltage survives, in contrast to conventional in-plane steady-state current generation. The out-of-plane electric polarization in 2D systems is useful for vertical stacking layers, yet they are less common in nature than in-plane polarized systems. The dominant photovoltage component in STO thin films arises from inter-band transitions between Ti-3*d* and O-2*p* orbitals, which is characterized by a pure real-space electron hopping under LPL, and its onset redshifts with increasing film thickness. The BPV response can be effectively tuned via film thickness, surface termination, orientations, i.e., (001) and (111), and Moiré twisting. In contrast, monolayer and bilayer $Sr_2TiO_4$ remain BPV inactive due to preserved bulk-like symmetry, but Moiré-type twisting induces (chiral) interlayer coupling and breaks inversion symmetry, enabling an out-of-plane photovoltage (rather than photocurrent) controlled by both lattice structure and CPL chirality. Our findings uncover a general mechanism by which nominally centrosymmetric layer-resolved materials acquire polar photovoltaic responses at the 2D limit, offering new opportunities for designing tunable, symmetry-driven photonic and optoelectronic functionalities in oxide nanomaterials.

## Computational Methods

***Density functional theory***: The DFT calculations are performed using the Vienna *ab initio* simulation package (VASP) [46-48], with the projector augmented-wave (PAW) method [49] in PSLIBRARY [50]. The exchange-correlation interaction is treated by the Perdew-Burke-Ernzerhof (PBE) functional [49], and a plane-wave basis set with an energy cutoff of 500 eV is used to expand the valence electrons. A vacuum space of over 15 Å along the *z* direction is added to eliminate the artificial layer interactions. The first Brillouin zone integration is sampled by Γ-centered Monkhorst−Pack $k$-point meshes [51] with a grid of 13 × 13 × 1. The van der Waals (vdW) interaction is described by the DFT-D3 method [52]. The spin-orbit coupling (SOC) is included self-consistently throughout all the calculations. We use the WANNIER90 code [53,54] to construct a tight-binding model of the electronic Hamiltonian. The DFT-derived Bloch bands were projected onto a localized Wannier basis comprising all Ti-3d and O-2p orbitals.

***Band theory of the bulk photovoltaic effect***: The conventional BPV effect involves the generation of both shift current and injection current. Under alternating electric field $E$ at frequency $\omega$, the LPL and CPL induce shift current $[\sigma^{abc}(0;\omega,-\omega)]$ and injection current $[\eta^{abc}(0;\omega,-\omega)]$ photoconductances respectively. The $\sigma^{abc}$ and $\eta^{abc}$ are the photoconductivity tensors, namely, the induced current density is $j^a=\sigma^{abc}\mathrm{Re}[E_b^{\omega}E_c^{-\omega}]+\eta^{abc}\mathrm{Im}[E_b^{\omega}E_c^{-\omega}]$. Here, *a*, *b*, *c* refer to Cartesian coordinates in the *xy* plane (for 2D systems), and *a* represents the current direction and *b*, *c* represent the light polarized direction. According to the second-order Kubo perturbation theory, in the velocity gauge framework, they can be evaluated via [15]

$$\chi^{abc}=\frac{e^3}{(\hbar\omega)^2}\int\frac{d^3\boldsymbol{k}}{(2\pi)^3}\sum_{m,n,l}\frac{f_{lm}v_{lm}^b}{E_{ml}-\hbar\omega+i\delta}\left(\frac{v_{mn}^a v_{nl}^c}{E_{mn}+i\delta}-\frac{v_{mn}^c v_{nl}^a}{E_{nl}+i\delta}\right), \quad (1)$$

Here, $f_{lm}=f_l-f_m$ and $E_{ml}=E_m-E_l$ are band differences of Fermi-Dirac occupation and eigenenergy, respectively. $v_{lm}^b=\langle l|\hat{v}^b|m\rangle$ is the interband velocity matrix element along *b* direction, and $\delta$ measures carrier lifetime phenomenologically incorporating electron-electron correlation, electron-phonon scattering, extrinsic impurity and disorder effect. While in principle the carrier lifetime depends on the specific band index *n* and wavevector **k**, its precise computation remains highly challenging for realistic materials. Therefore, following the conventional approach, we adopt a universal phenomenological value of 0.2 ps throughout our calculations. We emphasize that this simplification does not compromise the main conclusions of this work. Our chosen value of 0.2 ps is conservative, as it is comparable with experimental and theoretical estimates reported in the literature, such as 0.4 ps for $CrI_3$ [55] and 0.2 ps for Ge [56,57]. Consequently, this choice ensures that the calculated injection photovoltaic magnitude is not overestimated. The (LPL) shift current photoconductance is then $\sigma^{abc} \equiv \frac{1}{2}\mathrm{Re}\{\chi^{abc} + \chi^{acb}\}$, and the injection current (CPL) photoconductance is $\eta^{abc} \equiv \frac{1}{2}\mathrm{Im}\{\chi^{abc} - \chi^{acb}\}$.

In the current work, we will focus on another type of BPV effect, in which the induced polarization is along the non-periodic direction (*z*-direcrion). Then, the photo-excited wave package shift along the out-of-plane direction would result in an electric polarization instead of an electric current, and their corresponding susceptibility functions become [40,58]

$$\tilde{\chi}^{zbc} = \frac{e^2}{(\hbar\omega)^2}\int\frac{d^3\boldsymbol{k}}{(2\pi)^3}\sum_{m,n,l}\frac{f_{lm}v_{lm}^b}{E_{ml}-\hbar\omega+i\delta}\left(\frac{d_{mn}^z v_{nl}^c}{E_{mn}+i\delta}-\frac{v_{mn}^c d_{nl}^z}{E_{nl}+i\delta}\right), \quad (2)$$

Here, $d_{ml}^z = e\langle m|\hat{d}^z|l\rangle$ is the out-of-plane dipole matrix elements, and $\hat{d}^z = \sum_l\sum_n z_l\,|\psi_{n,l}\rangle\langle\psi_{n,l}|$ is the out-of plane position operator, with $z_l$ the *z*-coordinate of the orbital localized on the *l*th layer. This approach has been widely adopted in previous works treating the non-periodic direction [58-60]. Therefore, one can determine the shift-like and injection-like responses under LPL ($\tilde{\sigma}^{zbc} \doteq \frac{1}{2}\mathrm{Re}\{\tilde{\chi}^{zbc} + \tilde{\chi}^{zcb}\}$) and CPL ($\tilde{\eta}^{zbc} \doteq \frac{1}{2}\mathrm{Im}\{\tilde{\chi}^{zbc} - \tilde{\chi}^{zcb}\}$). In the above equations, integrals in the whole Brillouin zone (BZ) are performed. In three-dimensional (3D) periodic boundary conditions, this corresponds to dividing the *k*-weighted integrands by the unit cell volume, which contains the lattice constant along *z* (denoted as $L_z$). Note that here the susceptibility of photo-induced polarization is defined within a three-dimensional system and therefore inherently depends on the supercell volume, which includes the artificial vacuum space contribution. To provide susceptibility values in the conventional 3D unit system but eliminate the vacuum effect, we multiply a factor of $\frac{L_z}{d_{\mathrm{eff}}}$, where $d_{\mathrm{eff}}$ refers to the effective thickness of STO thin films, which are taken to be 2.03 Å, 5.80 Å, and 9.96 Å for $Sr_xTi_xO_{3x}$ (*x*=1-3), respectively, and 11.46 Å for twisted bilayer $Sr_2Ti_1O_4$. This approach follows the established procedure for computing the BPV effect in 2D materials [61,62].

## Results

***Crystal structure***: The bulk perovskite STO adopts a cubic phase with centrosymmetric *Pm*3*m* symmetry (space group no. 221). DFT optimizations yield a lattice parameter of 3.89 Å, in excellent agreement with the experimental value of 3.90 Å [63]. In $ABO_3$-type perovskites, two distinct surface terminations are experimentally achievable for both $A_{x+1}B_xO_{3x+1}$ and its infinite counterparts $ABO_3$: the AO and $BO_2$ termination [64]. For (001)-oriented STO thin films, asymmetric $TiO_2$ and SrO terminations on the two sides lift the inversion symmetry of the cubic lattice in the bulk state, giving rise to a polar structure with *P*4*mm* symmetry (space group No. 99) (see Figure 1a). In this study, we refer to their thickness to the unit cell number along *z*, i.e., one unit cell (1 u.c.), two unit cells (2 u.c.), and three unit cells (3 u.c.) configurations as $Sr_xTi_xO_{3x}$ ($x$ = 1–3). The optimized in-plane lattice parameters are 3.83 Å, 3.85 Å, and 3.86 Å for $x$ = 1, 2, and 3, respectively. The corresponding electronic band structures of different structures of 2D $SrTiO_3$ systems [Figure 1(a,b)] and optical absorptions are provided in the Supplemental Material (SM) [65], Figure S1-S4 in Section I and II.

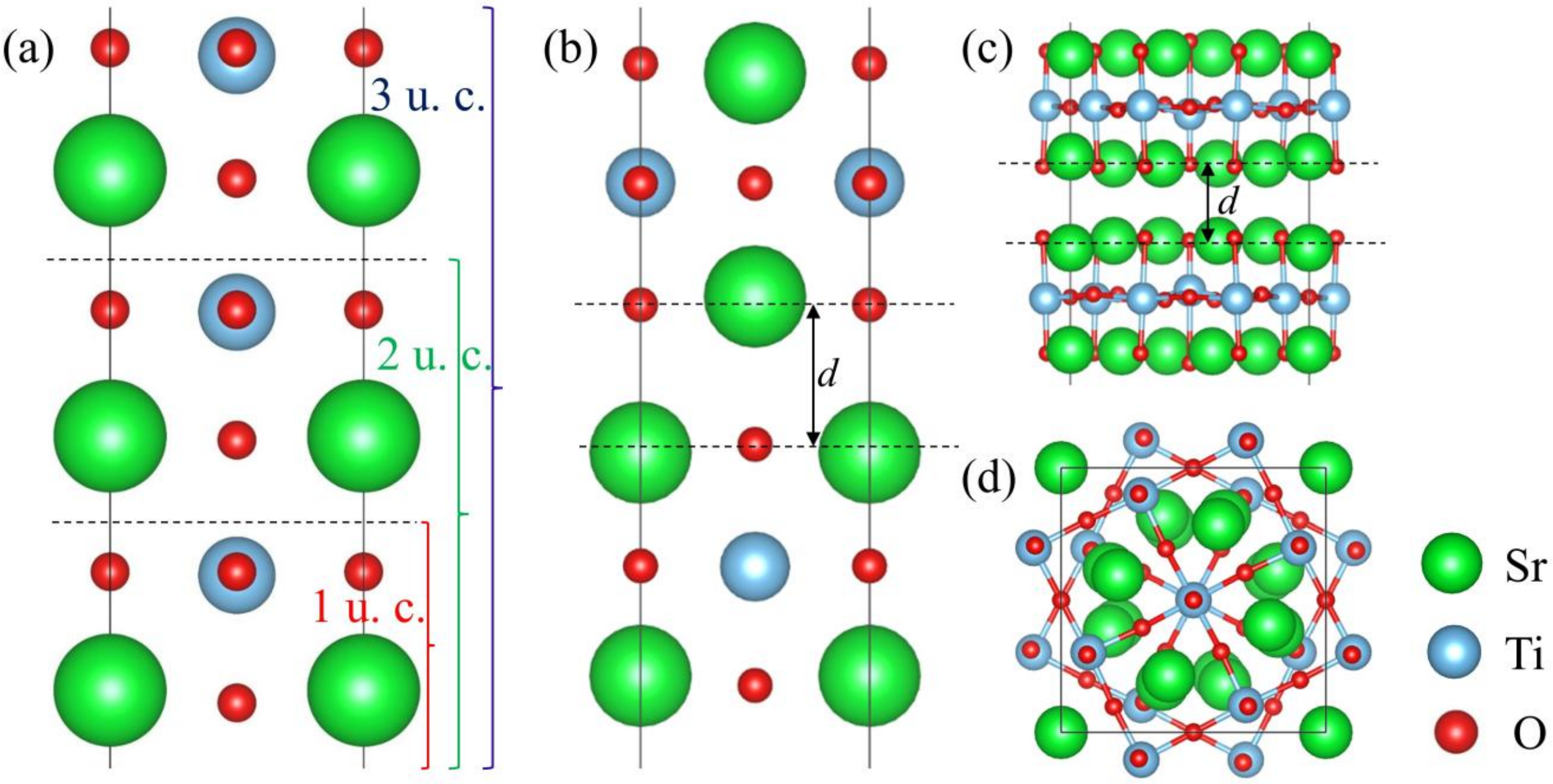

Figure 1. (a) Side view of $Sr_xTi_xO_{3x}$ ($x$ = 1-3) thin films and (b) bilayer $Sr_2Ti_1O_4$. (c, d) side and top views of twisted bilayer $Sr_2Ti_1O_4$ with twisting angle 36.87̊.

For the SrO termination on both sides of the finite-thickness thin films, the structure can be described by the formula $Sr_{x+1}Ti_xO_{3x+1}$, corresponding to Ruddlesden-Popper-type (RP) phase. For multilayer $Sr_{x+1}Ti_xO_{3x+1}$ superlattices, the SrO–SrO interface between two $Sr_{x+1}Ti_xO_{3x+1}$ films exhibits a van der Waals (vdW)-like interlayer interaction. Figure 1b shows the most energetically favorable AB-stacked bilayer $Sr_2Ti_1O_4$, which preserves inversion symmetry. The optimized lattice constant is 3.84 Å, and the interlayer distance is $d$ = 2.66 Å. The corresponding band structures for the monolayer and bilayer $Sr_2Ti_1O_4$ are provided in Section I and Figure S3 in SM [65].

A Moiré-twisted bilayer configuration can break the inversion symmetry. Figures 1c-d display the top and side views of a twisted bilayer $Sr_2Ti_1O_4$, in which the top layer is rotated clockwise relative to the bottom layer. The average interlayer distance of the twisted bilayer is $d$ = 2.75 Å—slightly larger than that of the AB-stacked configuration—consistent with trends observed in twisted $BaTiO_3$ (BTO) [30]. The symmetry of the system is reduced to the space group of *P*422 (No. 89).

Here, we take the 36.87° twisted configuration as a representative example, as it corresponds to the smallest commensurate supercell with a lattice constant of 8.48 Å. Further band structure and BPV calculations are provided in Section III and Figures S5–S6 [65].

The thin films discussed above are at the (001) plane of bulk STO. Other 2D counterparts with different orientations, such as the (111) plane, are also experimentally achievable. These exhibit distinct electronic structures and can support in-plane shift currents as well as out-of-plane photovoltages under LPL. For a detailed discussion of (111)-oriented STO films, see Section IV and Figure S7 [65].

***Symmetry analysis on BPV***: The lack of centrosymmetry in thin films of STO could give birth to the second NLO effect, such as the BPV effect. Before performing numerical calculations, we conduct a group symmetry analysis on the BPV generation. As both voltage and current transform as polar vectors, the symmetry arguments are the same for BPV current (along the thin film plane) and voltage (out-of-plane direction) generations [40].

As the intrinsic $Sr_xTi_xO_{3x}$ ($x$ = 1-3) thin films belong to the *P*4*mm* layer group without inversion symmetry, the LPL and CPL responses can be analyzed by investigating the $C_{4v}$ point group. The normal incident LPL and CPL follow the symmetric second order electric field representation, $\Gamma^{s}_{\boldsymbol{EE}^*} = A_1 \oplus B_1 \oplus B_2$ and asymmetric second order electric field representation, $\Gamma^{a}_{\boldsymbol{EE}^*} = A_2$. The in-plane LPL-current follows $\Gamma_{j_{x,y}} = E \otimes (A_1 \oplus B_1 \oplus B_2) = 3E$, giving no identical representation. Hence, they are completely forbidden. For the out-of-plane voltage, it obeys $\Gamma_{D_z} = A_1 \otimes (A_1 \oplus B_1 \oplus B_2) = A_1 \oplus B_1 \oplus B_2$, giving one independent component $\tilde{\sigma}^{zyy} = \tilde{\sigma}^{zxx}$. Thus, LPL would generate a BPV response along the *z* direction.

For the monolayer and bilayer $Sr_2Ti_1O_4$ (without Moiré twist), BPV is forbidden due to the inversion symmetry. As for the twisted bilayer $Sr_2Ti_1O_4$ without inversion symmetry, the BPV effect could be analyzed through the $D_{4h}$ symmetry. Normal incident LPL and CPL follow the symmetric and antisymmetric second order electric field representations, respectively, $\Gamma^{s}_{\boldsymbol{EE}^*} = A_1 \oplus B_1 \oplus B_2$ and $\Gamma^{a}_{\boldsymbol{EE}^*} = A_2$. The in-plane and out-of-plane LPL-induced BPV follows $\Gamma_{j_{x,y}} = E \otimes (A_1 \oplus B_1 \oplus B_2) = 3E$ and $\Gamma_{D_z} = A_2 \otimes (A_1 \oplus B_1 \oplus B_2) = A_2 \oplus B_2 \oplus B_1$ respectively, giving no symmetrically allowed components. For the CPL irradiation, the in-plane and out-of plane responses follow $\Gamma_{j_{x,y}} = E \otimes A_2 = E$ and $\Gamma_{D_z} = A_2 \otimes A_2 = A_1$. Thus, the CPL could result in one independent nonzero voltage susceptibility along the *z* direction, namely, $\tilde{\eta}^{zxy} = -\tilde{\eta}^{zyx}$.

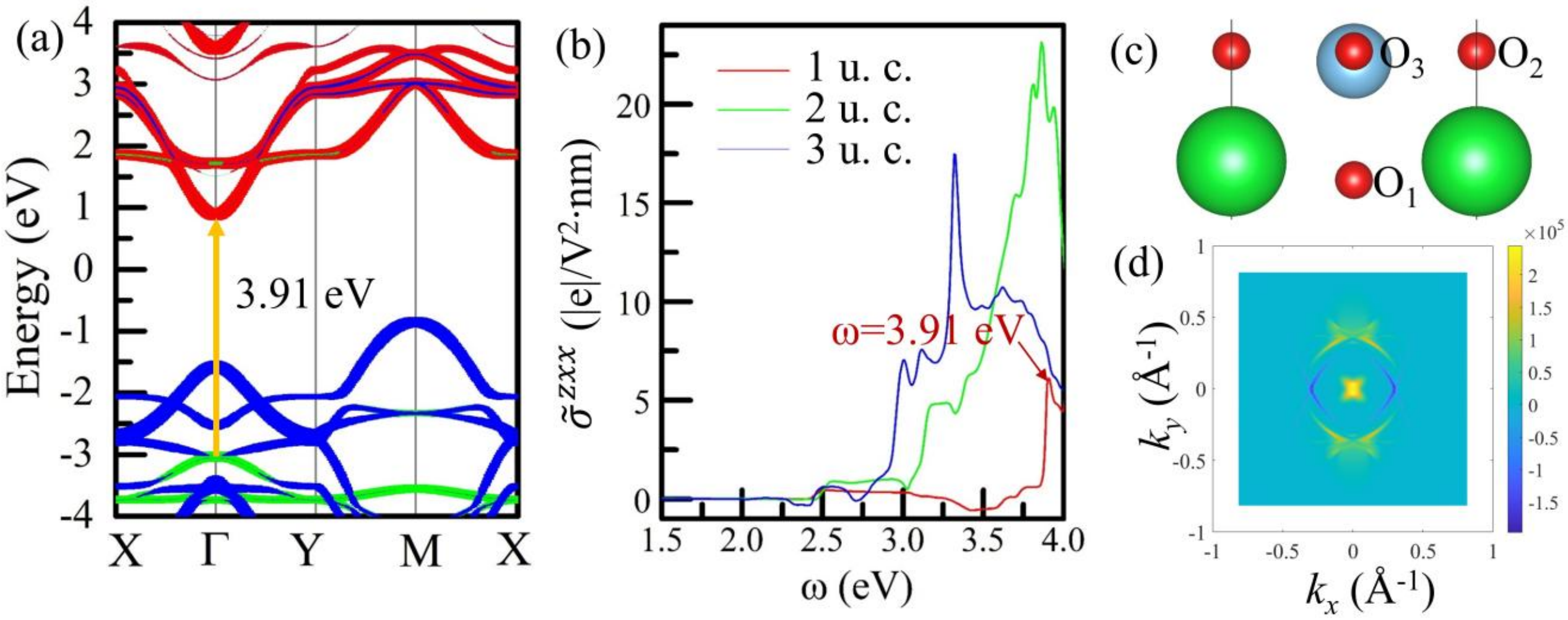

Figure 2. (a) Orbital-resolved band structure of 1 u. c. $SrTiO_3$. The red, blue, and green dotted lines represent Ti-3$d$ orbitals, top O-2$p$ orbital ($O_2$ + $O_3$), and bottom O-2$p$ orbital ($O_1$), respectively, as shown in panel (c). (b) Out-of-plane BPV coefficient of $Sr_xTi_xO_{3x}$ ($x$ = 1–3) thin films under LPL. (c) Structural model of 1 u. c. $SrTiO_3$ showing O atoms located at two distinct $z$-coordinates. (d) $k$-resolved BPV coefficient integrand distribution for 1 u. c. $SrTiO_3$ at a photon energy of 3.91 eV.

***DFT results for BPV responses***: We further calculate the BPV in $Sr_xTi_xO_{3x}$ ($x$ = 1-3) thin films (Figure 2a,b). Our numerical calculations reveal the same symmetry arguments as the previous analysis, and only $\tilde{\sigma}^{zxx}(\omega)$ is plotted here. This scales with quantum geometry $g_{nm}^{xx} = r_{nm}^{x} r_{mn}^{x} = |r_{nm}^{x}|^2$ and the band center difference $d_{nn}^{z} - d_{mm}^{z}$, where $r_{nm}^{x} = \frac{v_{nm}^{x}}{i\omega_{nm}}$ is the interband position operator matrix element. Therefore, in addition to the general symmetry requirement, the photon excited transitions should occur between band centers with different $z$ coordinates. In detail, we note that the direct bandgap of the 1 u. c. $SrTiO_3$ is 2.44 eV by PBE functional method, as shown in Figure 2 and Figure S1 [65]. However, the band edge contribution for $\tilde{\sigma}^{zxx}$ is negligible, and the dominant peak of the $\tilde{\sigma}^{zxx}$ is at the photon energy of 3.91 eV. To explain its mechanism, we analyze the atomic configuration and the orbital-resolved band structure in Figure 2a, c. In the 1 u.c. $SrTiO_3$, three O atoms are located at two different heights ($z$ coordinate), dubbed $O_1$ and $O_2$+$O_3$, respectively, as shown in Figure 2c. The orbital-resolved band structure in Figure 2a shows that the CBM and VBM are contributed by the Ti-3$d$ and $O_{2,3}$-2$p$ orbitals at the same height (same $z$ coordinate), respectively. In this regard, the out-of-plane dipole matrix mismatch $\Delta d_{nm}^{z} = d_{nn}^{z} - d_{mm}^{z}$ under electron excitation between CBM and VBM greatly diminishes. We see that the first dominant peak of the $\tilde{\sigma}^{zxx}$ lies at $\omega$ = 3.91 eV, exactly the energy difference between CBM and green dotted bands at Γ, which is contributed by the electron transition between Ti-3$d$ and $O_1$-2$p$ orbitals (green dotted line in Figure 2a), crossing the thin film thickness. This scenario is consistent with the fact that the Wannier center evolution in real space mainly contributes to the shift current. Note that in length gauge, the shift current is evaluated by $\sigma_{bb}^{a}(\omega) =$ $\frac{\pi e^3}{\hbar^2}\int\frac{d^3k}{8\pi^3}\sum_{m,n} f_{m,n} R_{mn}^{a;b}|r_{mn}^{b}|^2\delta(\omega_{nm}-\omega)$, where light is polarized along the $b$ direction [18]. The $|r_{mn}^{b}|^2\delta(\omega_{nm}-\omega)$ evaluates absorption rate from band $m$ to band $n$ according to the Fermi's golden rule. Figure S4 demonstrates that a certain optical absorption in the band edge at Γ [65]. $R_{mn}^{ab}=\partial_a\phi_{mn}^{b}-A_{mm}^{a}+A_{nn}^{a}$ is the shift vector, and $\phi_{mn}^{b}$ is the phase of $r_{mn}^{b}=|r_{mn}^{b}|e^{i\phi_{mn}^{b}}$. $R_{mn}^{a,b}$ has the unit of length and can be physically interpreted as the position change of a wave-packet during its transition from band $m$ to band $n$ [11,62]. In the current out-of-plane BPV responses, the intraband Berry connection $A_{nn}^{z}$ can be simplified as the Wannier center location in the $z$ direction. Since it is along the nonperiodic direction, $R_{mn}^{z}$ reduces to $\Delta d_{nm}^{z}$. Thus, negligible electron transition between Ti-$d$ and $O_{2,3}$-$p$ orbitals contributes to the $\tilde{\sigma}^{zxx}(\omega)$. The $\boldsymbol{k}$-dependent BPV distribution in the reciprocal space in Figure 2d also shows that the main contribution of the first peak is from the Γ point, consistent with the previous discussion.

A key question is how the susceptibility of photo-induced polarization evolves with sample thickness. To this end, we calculated the 1 u c. to 3 u.c. cases and plotted their spectra in Figure 2b. As the layer thickness increases, the first peak of $\tilde{\sigma}^{zxx}$ exhibits a redshift, and additional resonant peaks emerge. This behavior can be attributed to the reduction of the bandgap with increasing thickness (1.74 eV, 0.73 eV, and 0.54 eV for 1 u. c., 2 u. c., and 3 u. c., respectively, see Figure S1), as well as the presence of more Ti and O atoms at different vertical positions, which enhance the BPV response. It should be noted that for thicker films approaching the bulk limit, the BPV effect diminishes and eventually vanishes. This thickness-dependent electronic structure evolution distinguishes $SrTiO_3$ (STO) from many other 2D materials commonly studied for photovoltaic applications. In typical 2D materials, such as transition metal dichalcogenides (e.g., $MoS_2$, $WSe_2$) or $Bi_2O_2X$, the bandgap also decreases with increasing thickness, but the underlying mechanism is primarily quantum confinement. For STO, however, the thickness-driven bandgap reduction (from 1.74 eV in 1 u. c. to 0.54 eV in 3 u. c.) is accompanied by a unique structural feature – Ti and O atoms at distinct vertical positions contribute cooperatively to the BPV response. As more layers are added, the number of Ti-O bonds along the out-of-plane direction increases, providing additional optical transition channels that manifest as new resonant peaks in the BPV spectra. Furthermore, unlike 2D ferroelectric materials such as $CuInP_2S_6$ [66], where the BPV effect originates primarily from out-of-plane polarization and exhibits a sharp drop in photocurrent density when the thickness exceeds the carrier mean free path (~40 nm). Here, STO offers a wider tunability range at the few-layer scale. The ability to continuously modulate both the bandgap and the number of active Ti-O layers by simply varying the thickness provides a straightforward yet effective knob for engineering the BPV response in STO-based optoelectronic devices.

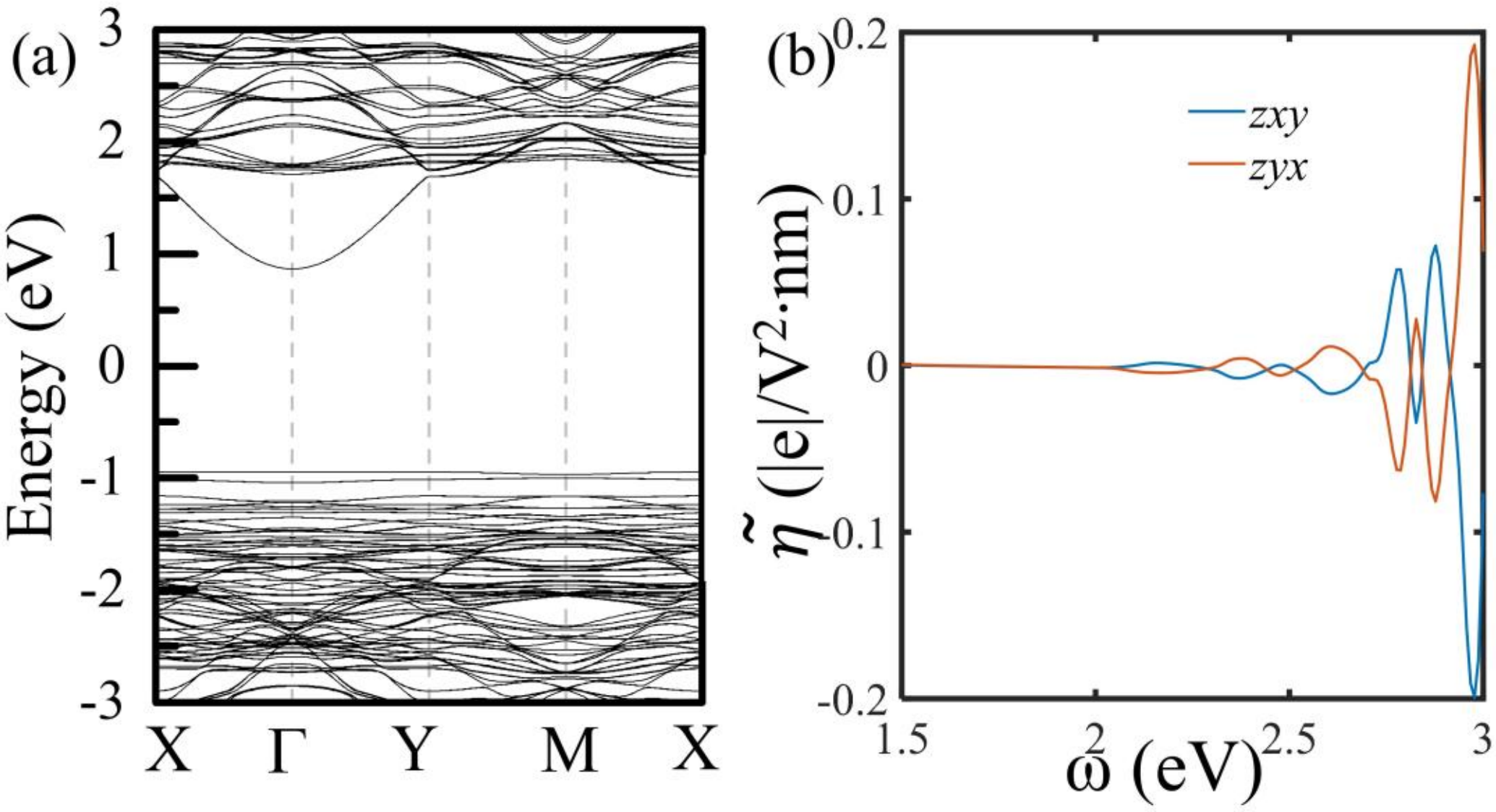

Figure 3. (a) Band structure and (b) out-of-plane BPV coefficient under CPL of Moiré twisted bilayer $Sr_2Ti_1O_4$.

For SrO terminations on both sides along the (001) orientation, the BPV effect is forbidden in monolayer and bilayer $Sr_2Ti_1O_4$ due to symmetry constraints. To induce a finite BPV response, we propose introducing mechanical shearing—specifically, a Moiré twist between the two layers—to break the symmetry. The band structure of the twisted bilayer $Sr_2Ti_1O_4$, shown in Figure 3a, exhibits a quasi-direct band gap of 1.81 eV at the Γ point. The symmetry-allowed out-of-plane susceptibility of photo-induced polarization is $\tilde{\eta}^{zxy}(\omega) = -\tilde{\eta}^{zyx}(\omega)$, as shown in Figure 3b, consistent with symmetry analysis. Since each $Sr_2Ti_1O_4$ layer is centrosymmetric, the emergence of $\tilde{\eta}^{zxy}(\omega)$ originates from interlayer electronic transitions, which must overcome the van der Waals (vdW) interactions between the layers.

From the projected orbital-resolved band structure shown in Figure S5 [65], the interband transitions between the lower O and upper Ti atoms, and vice versa, are identified as the primary contributors to the photovoltage. The $k$-resolved $\tilde{\eta}^{zxy}(\omega)$ integrand distributions for the main peaks are presented in Figure S6 [65], revealing clear symmetry-imposed patterns. The BPV coefficient $\tilde{\eta}^{zyx}$ at a photon energy of 3.0 eV is approximately 0.2 $|e|/V^2$. This indicates that under an incident electric field of $E$ = 0.1 V/nm, one yields an induced electric polarization of $d^z$ = 0.2 $|e|/Å^2$. This value is comparable to previously reported out-of-plane photovoltages in bilayer systems such as $WTe_2$ [58]. Notably, earlier studies of BPV effects in twisted structures have been limited to bilayer graphene, where the incident photon energy lies in the terahertz (THz) regime [36-40]. Moreover, the rotational angle—corresponding to the twist chirality—can also modulate the light-induced voltage, demonstrating a chiral photogalvanic effect. Note that while we are reporting photo-induced electric polarization susceptibility function under open circuit boundary condition, which describes the accumulated photocarriers at the top and bottom layers. In the steady-state, no current would exist along the out-of-plane direction. In order to measure the photocurrent, one could deposit both top and bottom electrodes on the STO samples and form a closed circuit condition, as in previous works for $CuInP_2S_6$ [66]. Additional appropriate 2D inert layers may be needed to preserve the intrinsic electronic properties of $SrTiO_3$.

Finally, to further explore the BPV responses, we examine the structures, electronic properties, and symmetry characteristics of 2D STO oriented along the (111) direction, as shown in Figure S7 in SM [65]. In contrast to the (001)-oriented STO, both in-plane shift current and out-of-plane photovoltage under LPL are allowed in this configuration. This result indicates that crystal orientation, alongside thickness, provides an additional degree of control.

Before concluding, we would like to make a few remarks. Firstly, STO here serves as a representative system, while analogous perovskites such as BTO [30] are expected to follow similar physical mechanisms. By tuning the hopping parameters in the tight-binding model, one can systematically model the electronic structure and extend the BPV response to a wide range of perovskite oxide systems (see detailed discussion in the SM [65]). In addition, we note that this work focuses primarily on the BPV generation in low-dimensional perovskites and its modulation. Accordingly, our DFT calculations are based upon a single-particle approximation and do not include quasiparticle or excitonic corrections, which are extremely computationally demanding. We remind the readers that both peak position and BPV magnitude calculated in the GGA-PBE functional may deviate from experimental measurements. As a thumb rule, one could expect that they primarily result in a redshift of the BPV spectral peak positions. Furthermore, as suggested by previous studies [67], many-body effects do not significantly alter the BPV effects in ultrathin 2D systems, as the light penetration depth is much larger than the film thickness. Hence, we report the DFT-based calculations, which do not affect the main physical picture and would provide comparable results with experimental observations, at least in the low-frequency regime. Moreover, it is worth noting that, although more advanced functionals such as Heyd-Scuseria-Ernzerhof (HSE) or Green's function and screened Coulomb interaction (GW) approximation would provide greater accuracy and result in a blue shift of the susceptibility of photo-induced polarization spectra, the structural complexity of the systems considered herein necessitates the use of the computationally more efficient PBE functional. Taking $Sr_xTi_xO_{3x}$ ($x$ = 1–3) thin films as an example, we present the HSE band structures in Figure S2 in SM [65] and the estimated band gaps are 3.29 eV, 2.32 eV, and 2.15eV for $x$ =1-3 respectively. We would like to note that the overall trend predicted using PBE should remain the same as in HSE or GW calculations, since there are no subtle topologically nontrivial band inversions or crossings.

## Conclusion

In conclusion, we systematically investigated the BPV responses in 2D STO with SrO and $TiO_2$ terminations along the (001) orientation. In these systems, the symmetry-allowed BPV components are $\tilde{\sigma}^{zxx} = \tilde{\sigma}^{zyy}$ for $SrTiO_3$ thin films under LPL and $\tilde{\eta}^{zxy} = -\tilde{\eta}^{zyx}$ for twisted bilayer $Sr_2Ti_1O_4$ under CPL. The primary contribution to $\tilde{\sigma}^{zxx}$ in $SrTiO_3$ arises from the mismatch of electron wave-packet centers along the out-of-plane direction. In contrast, the photovoltage in twisted bilayer $Sr_2Ti_1O_4$ originates from interlayer electronic transitions and reverses sign under opposite light or twist chirality. Additionally, 2D STO along the (111) orientation exhibits distinct BPV responses compared to the (001) orientation, highlighting the role of crystal anisotropy. Finally, we generalize our findings to a wide range of perovskites using a simplified tight-binding model. These theoretical insights may guide and stimulate further experimental studies of low-dimensional perovskite systems with diverse geometric configurations.

## Acknowledgements.

We acknowledge the support from the National Natural Science Foundation of China (NSFC) under Grant Nos. 12274342, 12422407, and 12374065. L. S. also acknowledges the support from the Key Research and Development Program of Shaanxi (2024QY2-GJHX-42). Additionally, the authors acknowledge support from Beijing PARATERA Technology Co., LTD for providing high-performance resources for contributing to the research results reported within this paper.

## Supplemental Material

# Bulk Photovoltaic Effect in Two-Dimensional Perovskite Oxides

Chunmei Zhang[1,2,3] Jian Zhou[4*], Liang Si[1,2,3*]

[1]School of Physics, Northwest University, Xi'an 710127, China
[2]Shaanxi Key Laboratory for Theoretical Physics Frontiers, Xi'an 710127, China
[3]Fundamental Discipline Research Center for Quantum Science and Technology of Shaanxi Province, Xi'an 710127, China
[4]Center for Alloy Innovation and Design, State Key Laboratory for Mechanical Behavior of Materials, Xi'an Jiaotong University, Xi'an, 710049, China

[*]Corresponding authors: jianzhou@xjtu.edu.cn; siliang@nwu.edu.cn

### I. Electronic structure of 2D $Sr_xTi_xO_{3x}$ ($x$=1-3) and $Sr_{x+1}Ti_xO_{3x+1}$ along (001) orientation

It should be noted that since the orbital contribution near the Fermi level predominantly originates from Ti 3*d* and O *p* states—both of which exhibit relatively weak spin orbital coupling (SOC)—the resulting band splitting is negligible. Furthermore, the Sr atoms mainly compose *s* orbitals (with small contributions of *d* orbitals), which do not give observable SOC (L = 0).

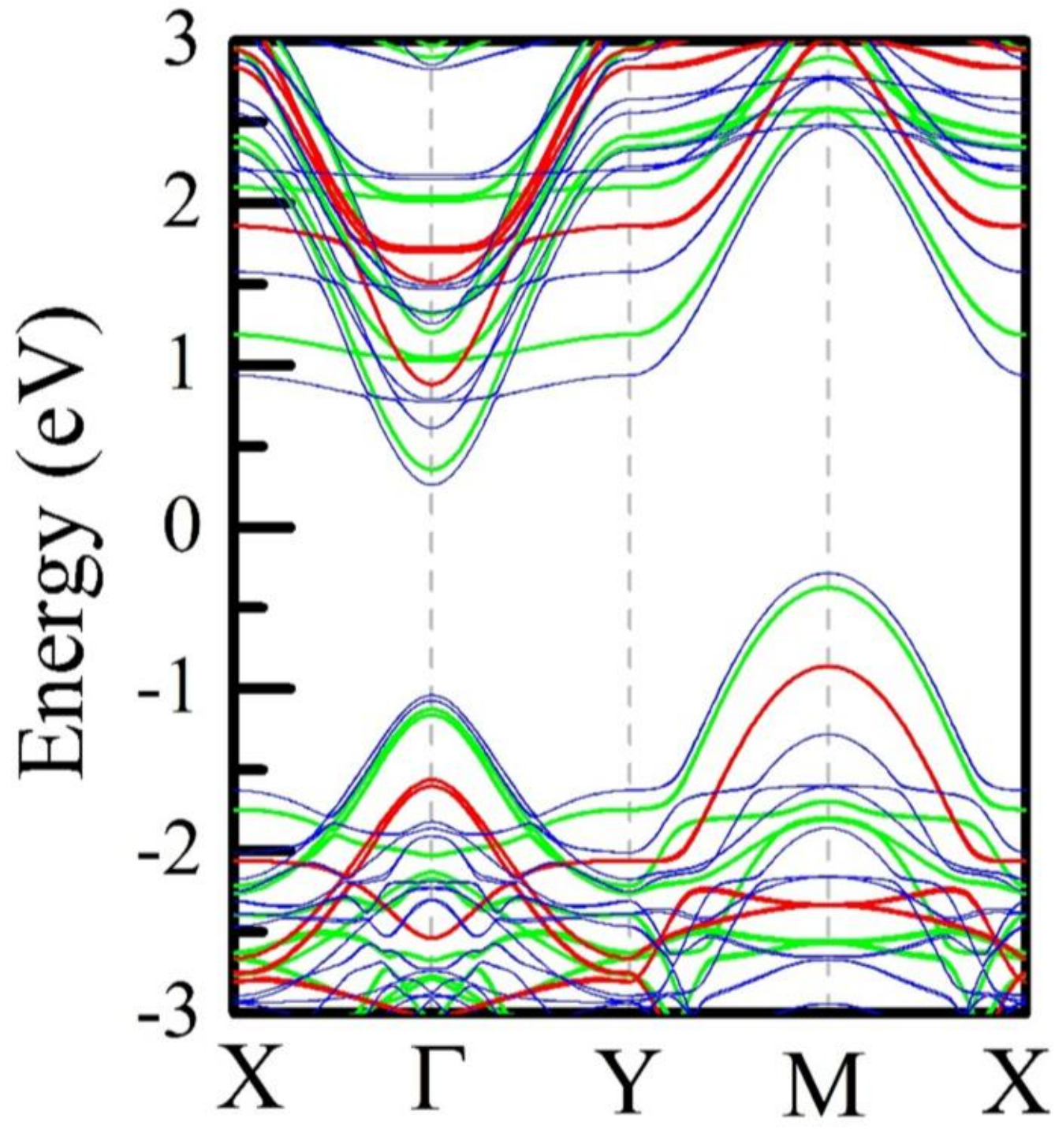

Figure S1. Calculated band structure of $Sr_xTi_xO_{3x}$ ($x$ = 1-3) thin films. The red, green, and blue lines represent one unit cell (1 u. c.), two unit cells (2 u. c.), and three unit cells (3 u. c.) $Sr_xTi_xO_{3x}$, respectively.

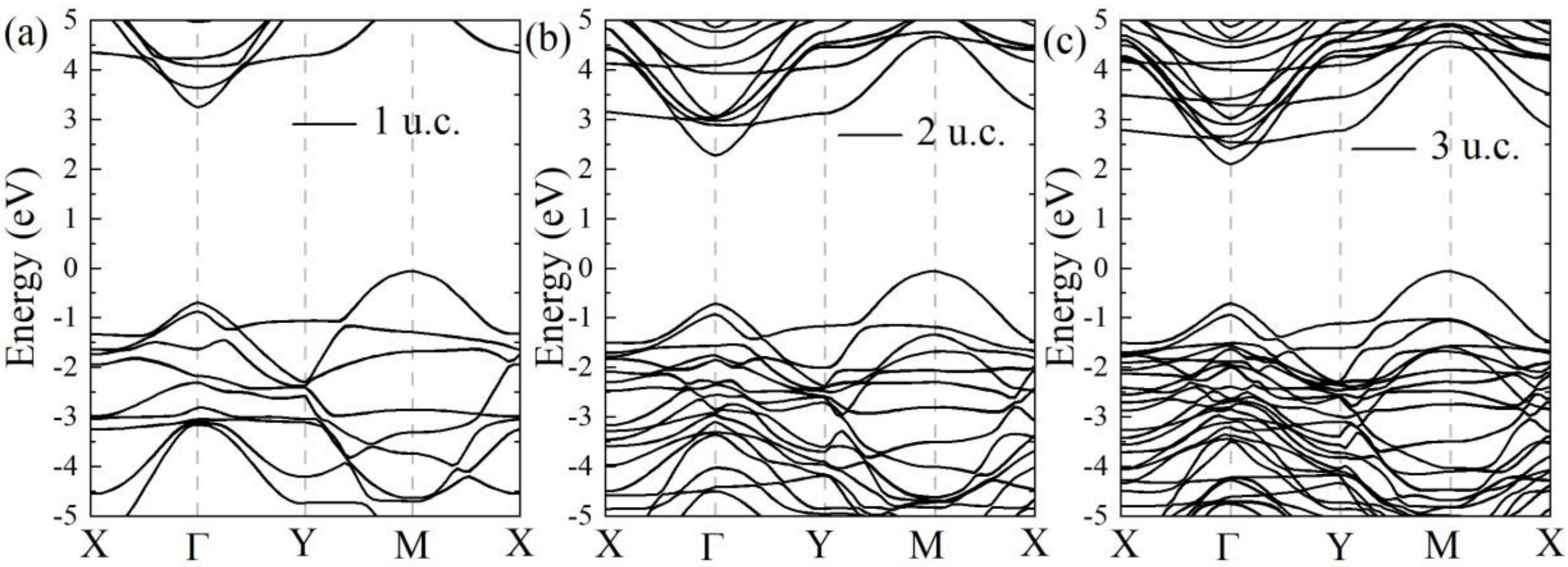

Figure S2. Calculated band structure of $Sr_xTi_xO_{3x}$ ($x$ = 1-3) thin films, including full SOC effect by HSE functional method.

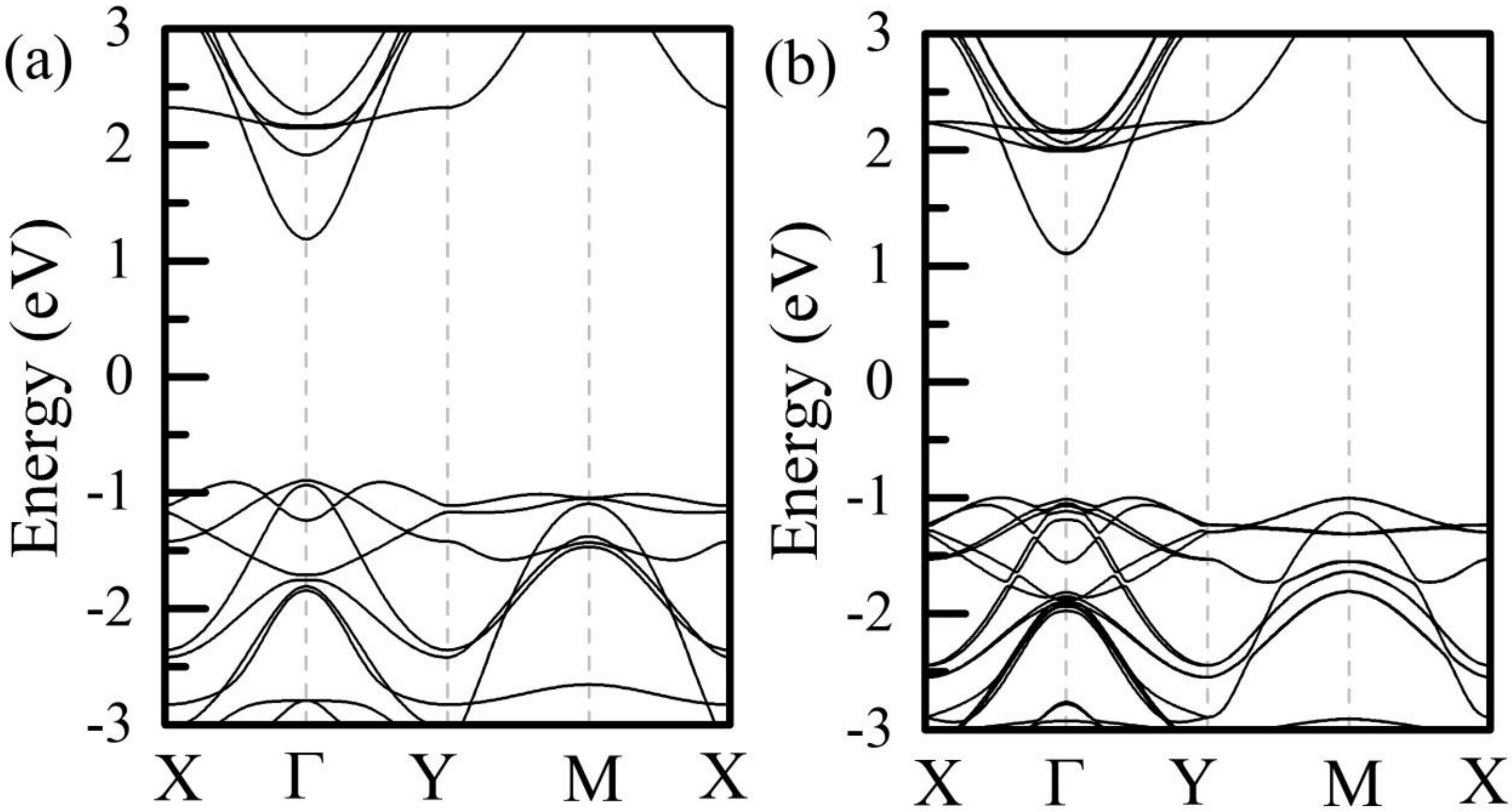

Figure S3. Band structures for (a) monolayer and (b) bilayer $Sr_2Ti_1O_4$ including full SOC effect by PBE functional method.

**II. Optical absorption of monolayer $SrTiO_3$ along (001) orientation.**

The imaginary part of the dielectric function of monolayer $SrTiO_3$, which represents the optical absorption, is presented in Figure S4. A large optical absorption is seen at the band edge at the **Γ** point.

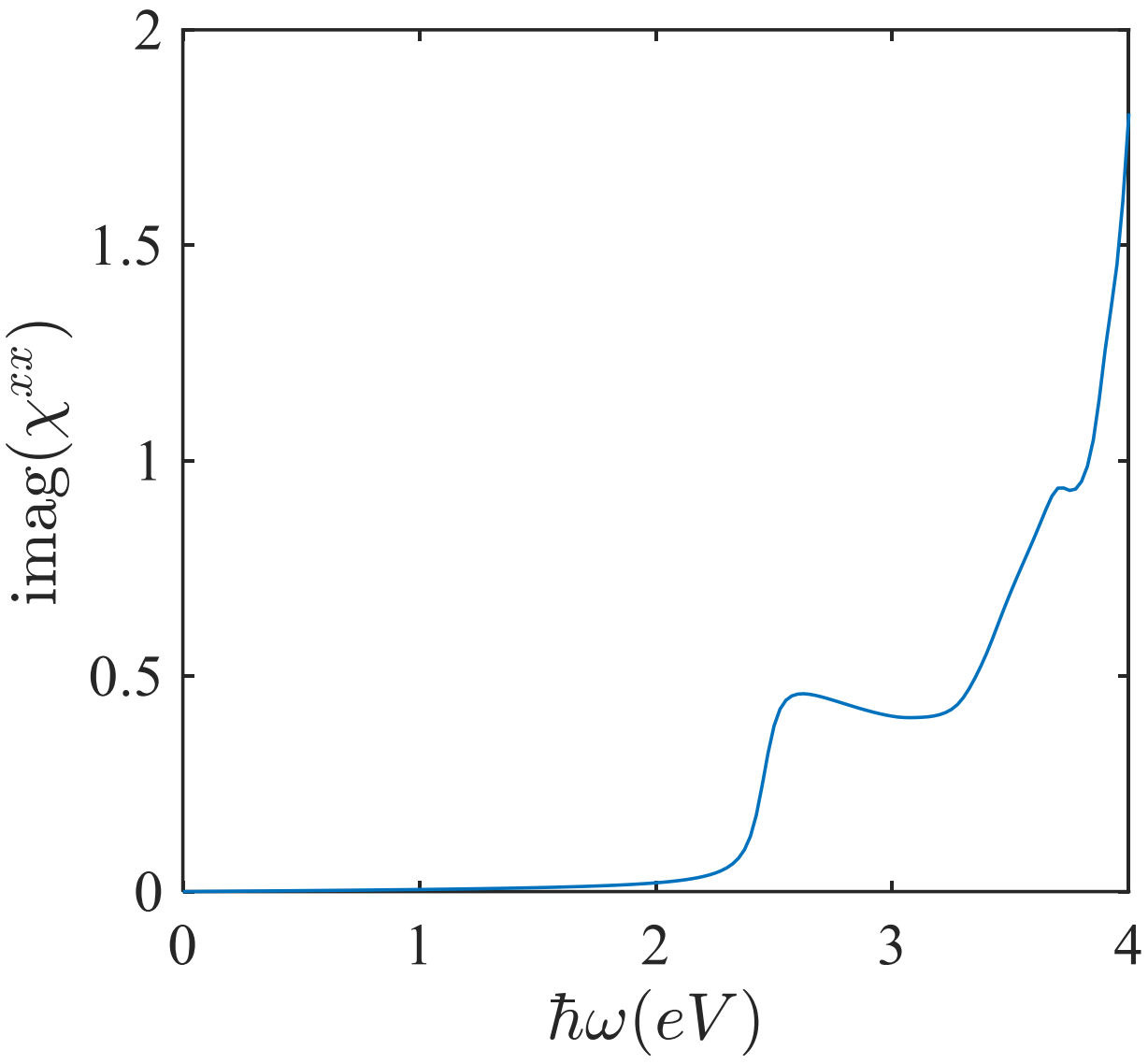

Figure S4, the imaginary part of the dielectric function of 1 u.c. $SrTiO_3$ along (001) orientation.

**III. Orbital-resolved band structures and *k*-resolved out-of-plane BPV coefficient in twisted bilayer $Sr_2Ti_1O_4$.**

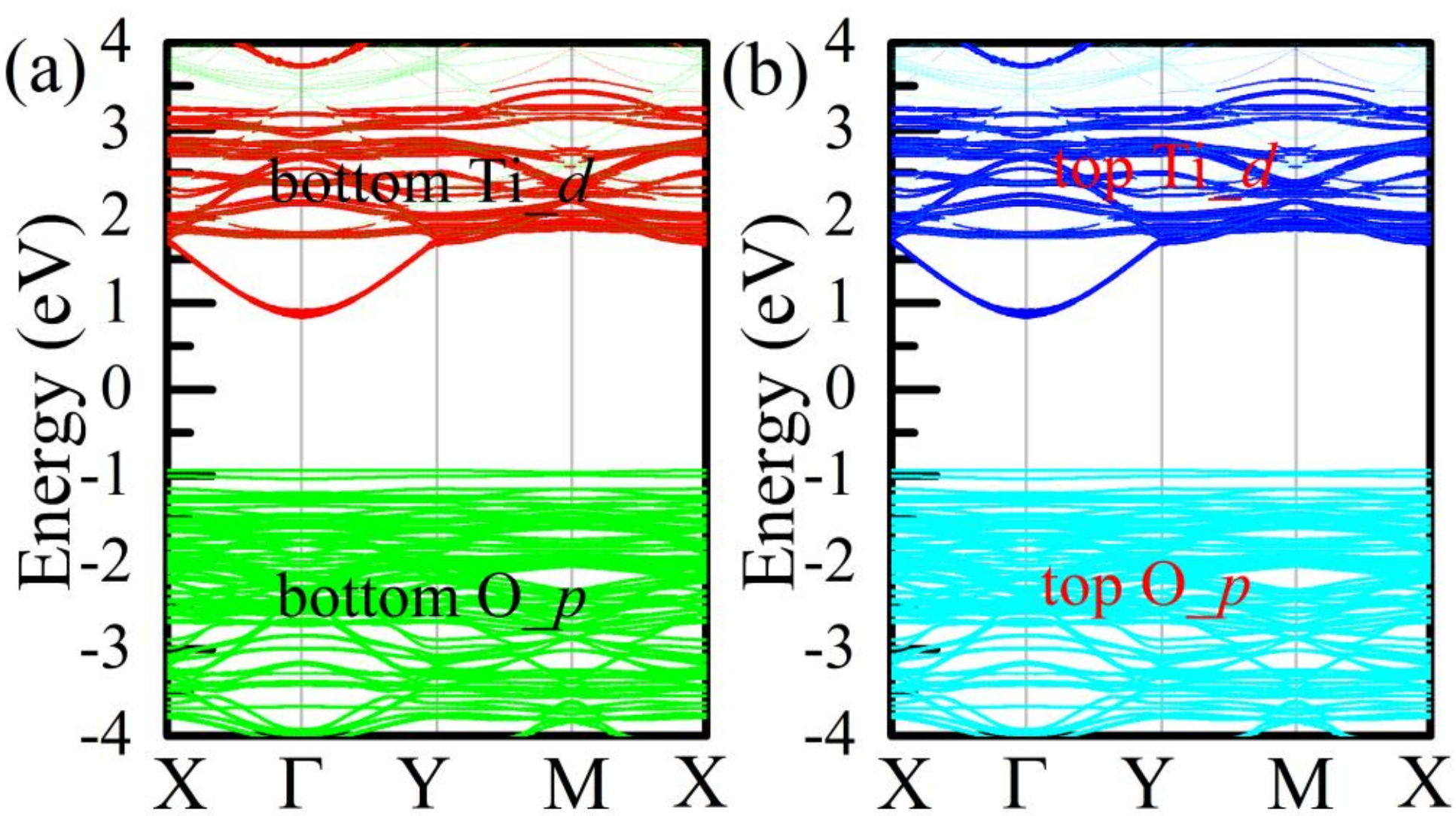

Figure S5. The calculated (a-b) orbital resolved band structure for twisted bilayer $Sr_2Ti_1O_4$. The red, green, blue, and cyan dotted lines represent the bottom and top Ti-*d* and O-*p* orbitals, respectively.

The photovoltage $\tilde{\eta}^{zxy}$ for twisted bilayer $Sr_2Ti_1O_4$ is presented in Figure S6a with the photovoltage peaks marked. The corresponding photovoltage distribution for every peak is given in Figure S6b-i.

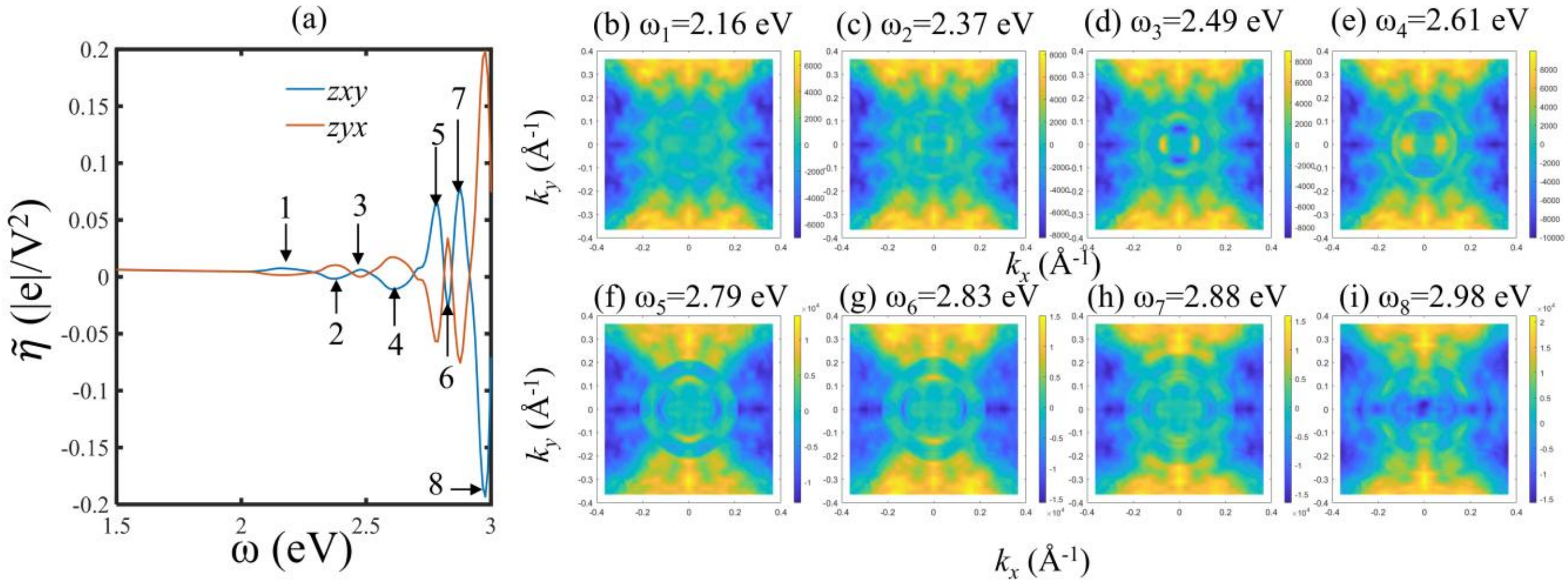

Figure S6. (a) The calculated out-of-plane BPV coefficient for twisted bilayer $Sr_2Ti_1O_4$ under CPL with all peaks of $\tilde{\eta}^{zxy}$ marked by black arrows. The calculated (b-i) the $k$-resolved BPV coefficient $\tilde{\eta}^{zxy}$ distribution for twisted bilayer $Sr_2Ti_1O_4$ at photon energy marked in Figure a.

**IV. Electronic structure and BPV of 2D (111) orientation of $SrTiO_3$.**

When STO is exfoliated along the (111) orientation, the structures are hexagonal and belong to $C_{3v}$ point group. Figure S7a-b presents its monolayer and bilayer forms, and their lattice parameters are 5.36 Å and 5.52 Å, respectively. Their direct band gaps at **Γ** point are 1.072 eV and 0.271 eV, respectively.

The LPL and CPL induced shift current and injection current in this system can be understood by investigating the $C_{3v}$ point group character. We concentrate on the BPV response under LPL, which follows the symmetric second order electric field representation, $\Gamma^{s}_{\boldsymbol{EE}^*} = A_1 \oplus E$. The in-plane shift current generation follows $\Gamma = E \otimes (A_1 \oplus E) = A_1 \oplus E$, giving one independent component $\sigma^{yyy} = -\sigma^{yxx}$, while the out-of plane shift current generation obeys $\Gamma = A_1 \otimes (A_1 \oplus E) = A_1 \oplus E$, giving one independent component $\tilde{\sigma}^{zyy} = \tilde{\sigma}^{zxx}$. Thus, contrary to the structures along (001) orientation, the structures along (111) orientation allow for in-plane shift current and out-of-plane photovoltage under LPL.

Figure S7c-d illustrates the allowed independent BPV tensor components for monolayer and bilayer STO along (111) orientation, which are consistent with the symmetry analyses. It shows that the first peak of the $\sigma^{yyy}$ and $\tilde{\sigma}^{zyy}$ for bilayer redshift compared with that of monolayer, as the bandgap is smaller than that of monolayer. Note that the effective thickness for monolayer and bilayer STO along (111) orientation is 1.193 Å and 3.533Å, respectively.

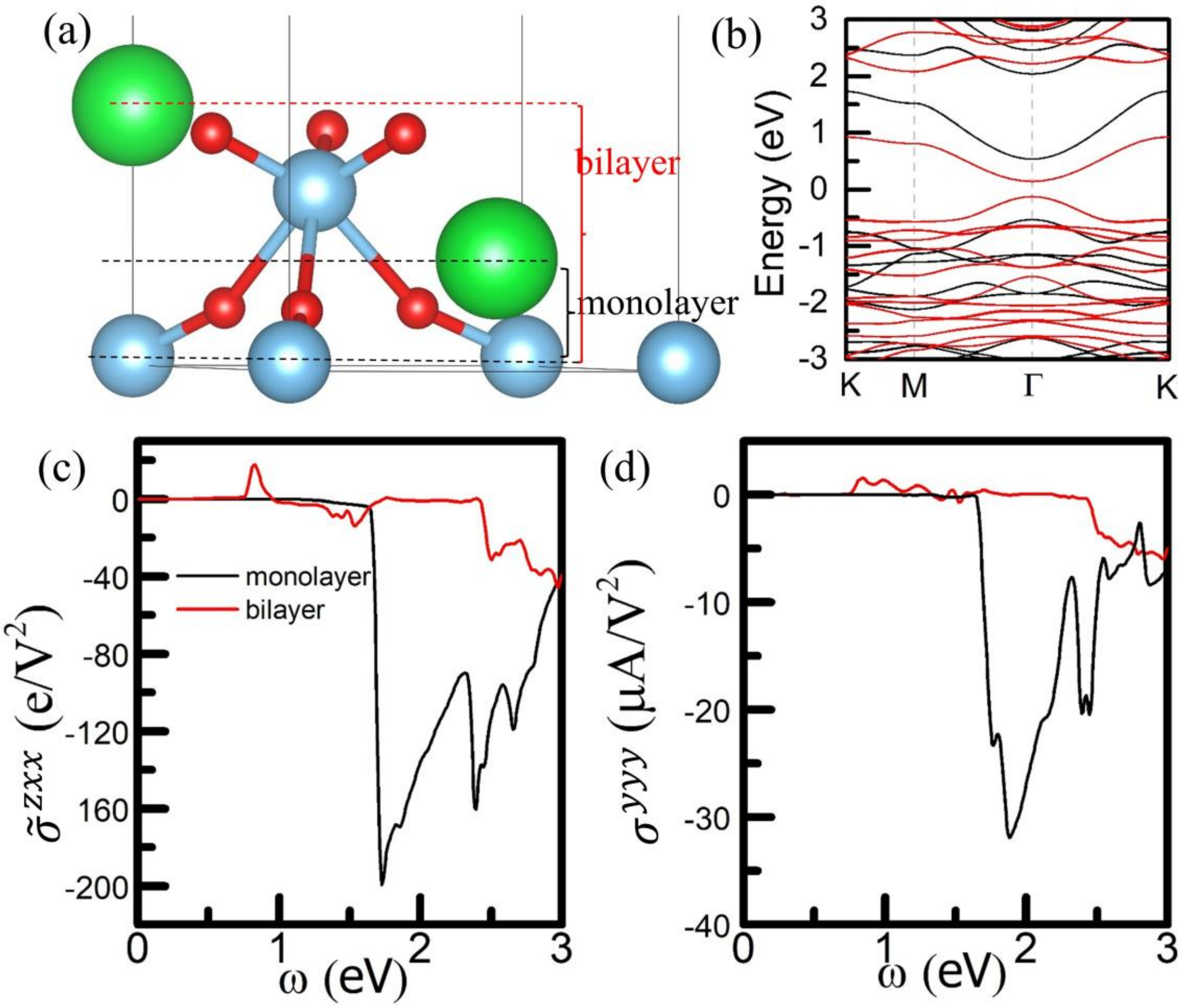

Figure S7. (a) The structure of $SrTiO_3$ along the (111) orientation with monolayer and bilayer forms. (b) The band structure of monolayer and bilayer $SrTiO_3$ along the (111) orientation. (c-d) The BPV coefficients under LPL $\tilde{\sigma}^{zxx}$ and $\sigma^{yyy}$ of monolayer and bilayer $SrTiO_3$ along the (111) orientation, respectively.

**V. Extension to other perovskite oxide systems:**

To gain microscopic insight into the low-energy electronic structure of ultrathin STO films with different orientations and to establish a simplified framework for describing BPV responses in perovskite oxides, we extracted the dominant orbital contributions and main hopping terms from maximally localized Wannier functions (MLWFs). The resulting Hamiltonian in momentum space is expressed as $H(k) = \sum_{\alpha\beta} t_{\alpha\beta}(R) e^{ik_x \cdot R_x + ik_y \cdot R_y}$, where $t_{\alpha\beta}$ denotes intra- and inter-orbital hopping integrals between orbitals $\alpha$ and $\beta$ within the Ti-3$d$ and O-2$p$ manifolds, and $\mathbf{R}_{x,y}$ is the lattice vector connecting them. For example, the Hamiltonian element for the Ti-3$d_{xy}$ orbital is $\varepsilon_{xy} = t_0 + 2t_1 [\cos(k_x) + \cos(k_y)] + 4t_2 \cos(k_x)\cos(k_y) + \cdots$, with fitted $t_0$ (2.464 eV) is the on-site potential, $t_1$(-0.150 eV) and $t_2$ (-0.014 eV) are the first and second nearest hopping along *x/y* (100 or 010) and $xy$ (110) direction. These hopping parameters encode the hybridization strength between neighboring sites, with large Ti–O terms reflecting the strong covalency of the perovskite lattice. Including both nearest-neighbor and longer-range hopping is essential to accurately reproduce the subtle band dispersions obtained from DFT and to capture BPV responses across different film thicknesses and orientations. Here, STO serves as a representative system, while analogous perovskites such as BTO are expected to follow similar physical mechanisms. By tuning $t_{\alpha\beta}$ , one can systematically model the electronic structure and BPV response of a wide range of perovskite oxide systems.